\documentclass[fleqn,usenatbib]{mnras}
\usepackage{newtxtext,newtxmath}
\usepackage[T1]{fontenc}

\DeclareRobustCommand{\VAN}[3]{#2}
\let\VANthebibliography\thebibliography
\def\thebibliography{\DeclareRobustCommand{\VAN}[3]{##3}\VANthebibliography}

\usepackage{graphicx}	
\usepackage{amsmath}	
\usepackage{color}

\title{An improved dynamical Poisson equation solver for self-gravity}

\author[R. Maeda et al.]{
Ryunosuke Maeda,$^{1}$\thanks{E-mail: maeda.ryunosuke@astr.tohoku.ac.jp}
Tsuyoshi Inoue$^{2}$
and Shu-ichiro Inutsuka$^{3}$
\\
$^{1}$Astronomical Institute, Tohoku University, 6-3 Aramaki, Aoba, Sendai 980-8578, Japan\\
$^{2}$Department of Physics, Konan University, Okamoto 8-9-1, Kobe 658-8501, Japan\\
$^{3}$Department of Physics, Graduate School of Science, Nagoya University, Furo-cho, Chikusa-ku, Nagoya 464-8602, Japan
}

\date{Accepted XXX. Received YYY; in original form ZZZ}

\pubyear{}

\begin{document}
\label{firstpage}
\pagerange{\pageref{firstpage}--\pageref{lastpage}}
\maketitle

\begin{abstract}
Since self-gravity is crucial in the structure formation of the universe, many hydrodynamics simulations with the effect of self-gravity have been conducted.
The multigrid method is widely used as a solver for the Poisson equation of the self-gravity; however, the parallelization efficiency of the multigrid method becomes worse when we use a massively parallel computer, and it becomes inefficient with more than $10^4$ cores, even for highly tuned codes. To perform large-scale parallel simulations ($> 10^4$ cores), developing a new gravity solver with good parallelization efficiency is beneficial. In this article, we develop a new self-gravity solver using the telegraph equation with a damping coefficient, $\kappa$. Parallelization is much easier than the case of the elliptic Poisson equation since the telegraph equation is a hyperbolic partial differential equation. We analyze convergence tests of our telegraph equations solver and determine that the best non-dimensional damping coefficient of the telegraph equations is $\tilde{\kappa} \simeq 2.5$. 
We also show that our method can maintain high parallelization efficiency even for massively parallel computations due to the hyperbolic nature of the telegraphic equation by weak-scaling tests. 
If the time step of the calculation is determined by heating/cooling or chemical reactions, rather than the CFL condition, our method may provide the method for calculating self-gravity faster than other previously known methods such as the fast Fourier transform and multigrid iteration solvers because gravitational phase velocity determined by the CFL condition using these timescales is much larger than the fluid velocity plus sound speed.
\end{abstract}



\section{Introduction}
Since self-gravity is crucial in different aspects of the universe, including the formation of a large scale structure, star formation, and planet formation, many simulations with self-gravity have been conducted to date. 
With the development of computers, higher-resolution and more detailed simulations have been conducted recently.
Massive parallelization is crucial for large-scale numerical simulations to reduce computation time; thus, developing massively parallelizable code is crucial for large-scale simulations.

Various methods for solving the Poisson equation have been developed to date for the numerical calculation of self-gravity. 
Methods using Green’s function expansion \citep[e.g.,][]{1995CoPhC..89...45M} and the fast Fourier transform (FFT) \citep[e.g.,][]{1979JCoPh..32...24E,Hockney_1988,2001JCoPh.170..231G,Press2007,2019ApJ...870...43M,2019ApJS..241...24M} are examples of self-gravity solvers. 
The FFT method is commonly used, and effective solvers have been developed for several situations such as periodic boundary conditions and isolated systems. 
However, the FFT method is generally applicable only to simple boundary conditions, and more importantly, parallelization efficiency is known to be very limited for massive parallel computations \citep{tomida2023athena++}.

The multigrid method \citep{brandt1977multi,press1986numerical} is another widely employed as the Poisson equation solver \citep[see also,][]{tomida2023athena++}. 
The multigrid method uses grids of different resolutions to quickly converge errors on long-wavelength scales, resulting in considerably faster convergence speeds compared to standard iterative methods (Jacobi method, Gauss-Seidel method, etc.).
However, the multigrid method is not very efficient when we use a parallel supercomputer MPI \citep[especially in flat MPI, see, e.g.,][]{nakajima2012new,nakajima2014optimization} with more than $10^4$ cores even for highly tuned code since the Poisson equation is an elliptic partial differential equation. 
OpenMP/MPI Hybrid parallel computing suggests that the multigrid method has a high performance even with a larger number of cores ($<10^5$) due to the reduced amount of communications among MPI processes \citep{nakajima2012new,nakajima2014optimization}, but for this, a high-level tuning of the code is required.
Thus, a new gravity solver with good parallelization efficiency must be developed to perform large-scale massive parallel computations with $> 10^4$ cores.

\cite{hirai2016hyperbolic} developed a new self-gravity solver by modifying the Poisson equation into an inhomogeneous wave equation, as in eq.~(\ref{eq:wv}). 
\begin{equation}
    \frac{\partial^2 \Phi}{\partial t^2}-c_{\mathrm{g}}^2 \nabla^2 \Phi=-4 \pi c_{\mathrm{g}}^2 G \rho, \label{eq:wv}
\end{equation}
where $c_\mathrm{g}$ represents the phase velocity of the gravitational wave, $\rho$ represents the density, and $G$ represents the gravitational constant.
Eq.~(\ref{eq:wv}) corresponds to the weak field limit of the Einstein equations in general relativity; therefore, the equation is physically suitable to use, and the solution is identical to that obtained by solving the Poisson equation when the gas velocity is negligibly small compared to the speed of light.
In \cite{hirai2016hyperbolic}, the computational error is estimated at $\Delta \sim (v^2+c_\mathrm{s}^2)/c_\mathrm{g}^2$, indicating that a sufficiently large $c_\mathrm{g}$ compared with the fluid velocity ($v$) and the sound speed ($c_\mathrm{s}$) must be employed in the simulation. 
Since the wave equation of gravity is a hyperbolic equation, it is ideal for the massively parallel computer. 
Additionally, their method is not expected to converge to the correct solution in the case of periodic boundary conditions. 

\cite{ruter2018hyperbolic} introduced a telegraph equation to solve the Poisson equation, which was solved by using a pseudospectral method. 
The pseudo-spectral method uses the spatial Fourier transform to solve the equations. 
They showed that the telegraph equation is effective for solving the Poisson equation, however, the FFT used in the pseudo-spectral method is generally only capable of simple boundary conditions such as periodic and isolated boundary conditions as mentioned before. 
Since the computational cost increases in complex problems where FFT cannot be used, it is important to develop a self-gravity solver using the telegraph equation that can be used in general problems. 
Therefore, in this study, we developed a self-gravity solver that directly calculates the telegraph equation. 
This requires solving the telegraph equation while avoiding the numerical instability of the hyperbolic equation. 
In addition, since \cite{ruter2018hyperbolic} used a specific dimensionless diffusion coefficient $\tilde{\kappa}=0.5$ ($\epsilon=1.0$, $\eta=1.0$ in their paper's expressions) of the telegraph equation, we investigate the optimal coefficient for convergence.
We also propose a method where we can solve the self-gravity with telegraph equation without any relaxation technique.

In this research, we further enhance the wave equation method by introducing the damping term of the wave that accelerates the convergence of the gravitational potential to the solution of the Poisson equation without spoiling the compatibility for the massive parallel computation.
This study is organized as follows.
In Section 2, we introduce the new set of equations to solve the self-gravity and give a setup for test simulations. 
In Section 3, we present the results of the simulations, and subsequently, in Sections 4 and 5, we discuss the results and summarize our study, respectively.

\section{method}
\subsection{Basic equations}
We consider the following telegraph-type equation for the gravitational potential $\Phi$:
\begin{equation}
    \frac{\partial^2 \Phi}{\partial t^2}+2 \kappa \frac{\partial \Phi}{\partial t}-c_{\mathrm{g}}^2 \nabla^2 \Phi=-4 \pi c_{\mathrm{g}}^2 G \rho, \label{eq:tel}
\end{equation}
where $\kappa$ represents the damping coefficient, $c_\mathrm{g}$ represents the phase velocity of the gravitational wave, $\rho$ represents the density, and $G$ represents the gravitational constant.
The steady solutions of eq.~(\ref{eq:tel}) correspond to those of the Poisson equation for self-gravity. 
Eq.~(\ref{eq:tel}) can be decomposed into the following first-order equations for $\Phi$ and, $\Psi$:
\begin{equation}
    \left( \frac{\partial }{\partial t} +c_{\mathrm{g}} \frac{\partial }{\partial x}+c_{\mathrm{g}} \frac{\partial }{\partial y}+c_{\mathrm{g}} \frac{\partial }{\partial z}+\kappa \right)\Phi=\kappa \Psi,
    \label{eq:a}
\end{equation}
\begin{eqnarray}
  & &\left( \frac{\partial }{\partial t} -c_{\mathrm{g}} \frac{\partial }{\partial x}-c_{\mathrm{g}} \frac{\partial }{\partial y}-c_{\mathrm{g}} \frac{\partial }{\partial z}+\kappa \right)\Psi=\kappa \Phi \nonumber \\
    & &-\frac{1}{\kappa} 4 \pi c_{\mathrm{g}}^{2} G \rho -\frac{1}{\kappa} 2 c_{\mathrm{g}}^{2} \left( \frac{\partial}{\partial x} \frac{\partial}{\partial y} + \frac{\partial}{\partial y} \frac{\partial}{\partial z} + \frac{\partial}{\partial z} \frac{\partial}{\partial x} \right) \Phi.
 \label{eq:b}
\end{eqnarray}
We can easily confirm that eqs.~(\ref{eq:a}) and (\ref{eq:b}) are identical to eq.~(\ref{eq:tel}) by eliminating $\Psi$ in eq.~(\ref{eq:b}) using eq.~(\ref{eq:a}).
Here we assume that $\kappa$ and $c_\mathrm{g}$ are constant values.

In this study, we solve eqs.~(\ref{eq:a}) and (\ref{eq:b}) by employing the operator-splitting technique.
The advection terms for the $\Phi$ and $\Psi$ are solved using the third-order MUSCL scheme.
The terms proportional to the damping coefficient $\kappa$ and the source term are solved using the piecewise exact solution method developed by \cite{inoue2008two}.
The second derivative terms of eq.~(\ref{eq:b}) are solved using the simple central difference explicit scheme.
The time steps $\Delta t$ are determined so that $\Delta t$ satisfies the Courant-Friedrichs-Lewy (CFL) conditions of $\Phi$ and $\Psi$, $\Delta t = 0.5 \Delta x / c_\mathrm{g}$.

\subsection{Simulation setups}

The non-dimensional form of eq.~(\ref{eq:tel}) can be written as
\begin{equation}
    \frac{\partial^{2} \tilde{\Phi}}{\partial \tilde{t}^{2}} +2 \tilde{\kappa} \frac{\partial \tilde{\Phi}}{\partial \tilde{t}} -\tilde{\nabla}^{2} \tilde{\Phi}=-\tilde{\rho}, \label{eq:non-d-tel}
\end{equation}
where $\tilde{t}=t c_\mathrm{g}/L$, $\tilde{x}=x/L,\ \tilde{y}=y/L,\ \tilde{z}=z/L,\ \tilde{\kappa}=\kappa L/c_\mathrm{g},\ \tilde{\Phi}=\Phi L/4 \pi G M,\ \textrm{and}\ \tilde{\rho}=\rho L^3/M$. Here, $M$ is the total gas mass in the numerical domain, and $L$ represents the size of the numerical domain.
Therefore, eq.~(\ref{eq:non-d-tel}) is characterized by the distribution of $\tilde{\rho}$ and $\tilde{\kappa}$. In what follows, we analyze how the solution of eq.~(\ref{eq:non-d-tel}) converges to the solution of the Poisson equation under the various choices of $\tilde{\rho}$ and $\tilde{\kappa}$.
For $\tilde{\kappa}$, we test the cases of $\tilde{\kappa}=5.0\times10^{-6},\ 5.0\times10^{-5},\ 0.5,\ 1.0,\ 1.667,\ 2.5,\ 3.333,\ 5.0,$ and $ 10.0$.
We will demonstrate that there is a certain value of $\tilde{\kappa}$ at which the dynamical solution of $\tilde{\Phi}$ most quickly converges to the solution of the Poisson equation.
In this research, we employ the same total mass in the numerical domain for all density distribution computations and employ a cubic numerical domain. 
We divide the numerical domain into $64^3, 128^3,$ and $256^3$ uniform cells to verify the convergence. 

When we use fixed boundary conditions (e.g., section \ref{sec:resl-statc}), we set the boundary conditions of $\tilde{\Phi}_{\mathrm{bd}}$ and $\tilde{\Psi}_{\mathrm{bd}}$ by using an analytic solution of the Poisson equation ($\tilde{\Phi}_{\mathrm{bd, exa}}$) as follows: 
\begin{equation}
    \tilde{\Phi}_{\mathrm{bd}}=\tilde{\Phi}_{\mathrm{bd,exa}}, \label{eq:bc-a}
\end{equation}
\begin{equation}
    \tilde{\Psi}_{\mathrm{bd}}=
\frac{1}{\tilde{\kappa}}\left( \frac{\partial}{\partial \tilde{x}}+ \frac{\partial}{\partial \tilde{y}}+\frac{\partial}{\partial \tilde{z}}+\tilde{\kappa}\right)\tilde{\Phi}_{\mathrm{bd,exa}}, \label{eq:bc-b}
\end{equation}
where eq.~(\ref{eq:bc-b}) is derived from eq.~(\ref{eq:a}) under the steady condition. 
Alternatively, when the boundary condition is set using a gravitational potential $\tilde{\Phi}_{\mathrm{bd,mlp}}$ obtained by multipole expansion or other methods, we may use  $\tilde{\Phi}_{\mathrm{bd,mlp}}$ instead of $\tilde{\Phi}_{\mathrm{bd,exa}}$ in eq.~(\ref{eq:bc-a}) and~(\ref{eq:bc-b}).

\section{result}
\subsection{Static density cases \label{sec:resl-statc}}

In this section, we check the gravitational potential’s convergence to the analytic solution of the Poisson equation when we start from $\tilde{\Phi}=0.0$. 
In addition, we analyze the dependence of the speed of convergence on $\tilde{\kappa}$. 
We define the difference between the analytic and numerical solutions as follows:
\begin{equation}
\Delta_\mathrm{ err }=\frac{\sqrt{\frac{1}{n_x n_y n_z} \Sigma_{i,j,k=1}^{n_x,n_y,n_z} \left(\tilde{\Phi}_{i,j,k}-\tilde{\Phi}_{\text {exa},i,j,k}\right)^{2}}}{\tilde{\Phi}_{\text {exa,max }}-\tilde{\Phi}_{\text {exa,min }}}, \label{eq:err}
\end{equation}
where $i, j,$ and $k$ denote a cell number in the $x, y,$ and $z$ directions, respectively; $n_x, x_y,$ and $n_z$ represent the total number of cells in the $x, y,$ and $z$ directions, respectively; $\tilde{\Phi}_\mathrm{exa}$ represents the analytic solution of the Poisson equation; and $\tilde{\Phi}_\mathrm{exa, max}$ and $\tilde{\Phi}_\mathrm{exa, min}$ represent the maximum and minimum value of the gravitational potential in the numerical domain, respectively. 
In what follows, we discuss the time evolution of $\Delta_\mathrm{ err }$ for different initial conditions.

\subsubsection{Case of a static uniform sphere}
\begin{figure*}
\includegraphics[width=1.0\textwidth]{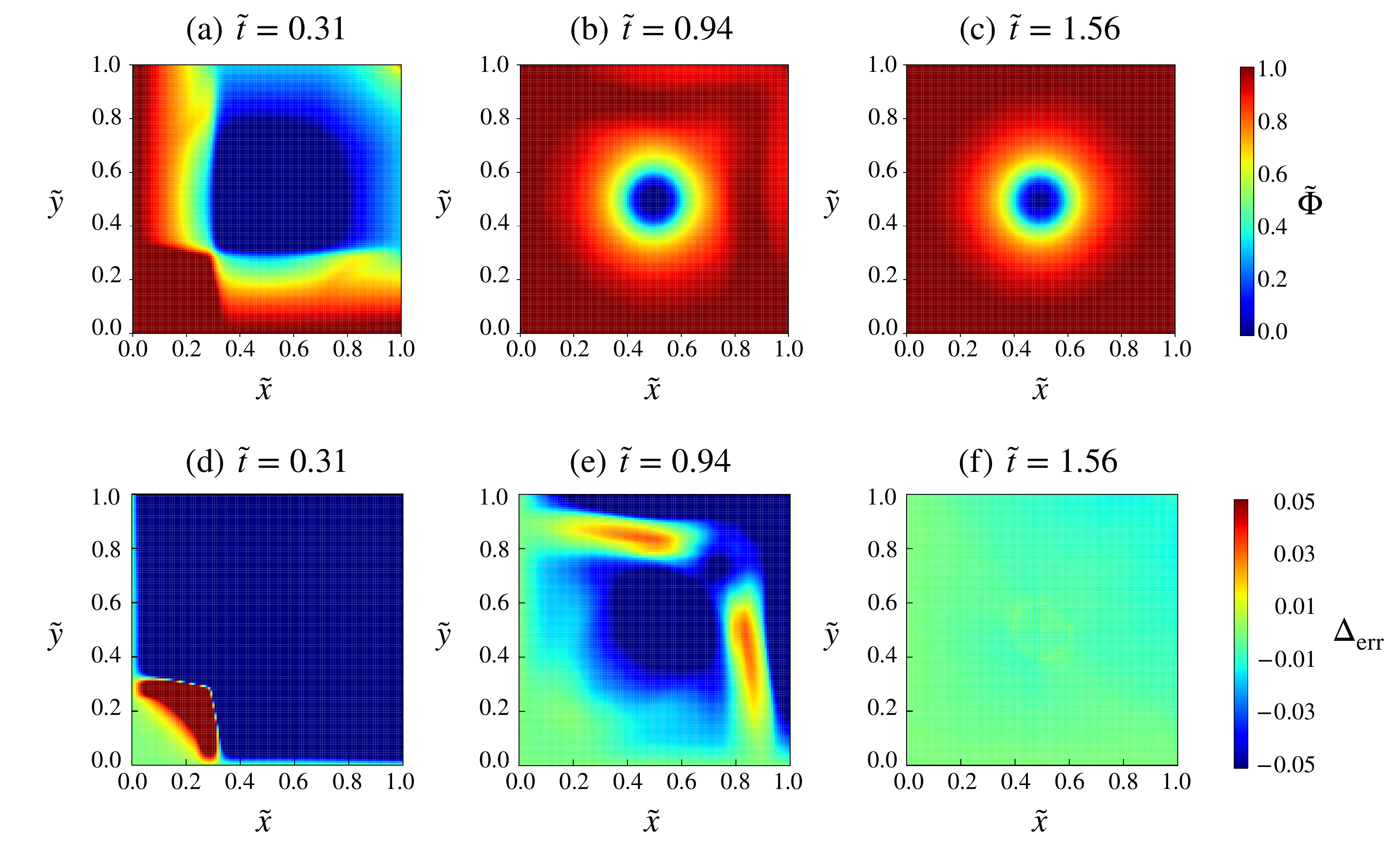}
\caption{Top panels indicate two-dimensional gravitational potential cross sections ($\tilde{z}=0.5$) of $\tilde{\kappa}=2.5$ results at (a) $\tilde{t}=0.31$, (b) $\tilde{t}=0.94$, and (c) $\tilde{t}=1.56$. 
The bottom panels indicate two-dimensional cross-sections of $\Delta_\mathrm{err}$ ($\tilde{z}=0.5$) at (d) $\tilde{t}=0.31$, (e) $\tilde{t}=0.94$, and (f) $\tilde{t}=1.56$. 
The initial density distribution is a uniform density sphere with $\tilde{r}=0.1$ and $\tilde{\rho}=10^3$.
 \label{fig:time-evl}}
\end{figure*}
\begin{figure}
 \centering
 \includegraphics[width=0.4\textwidth]{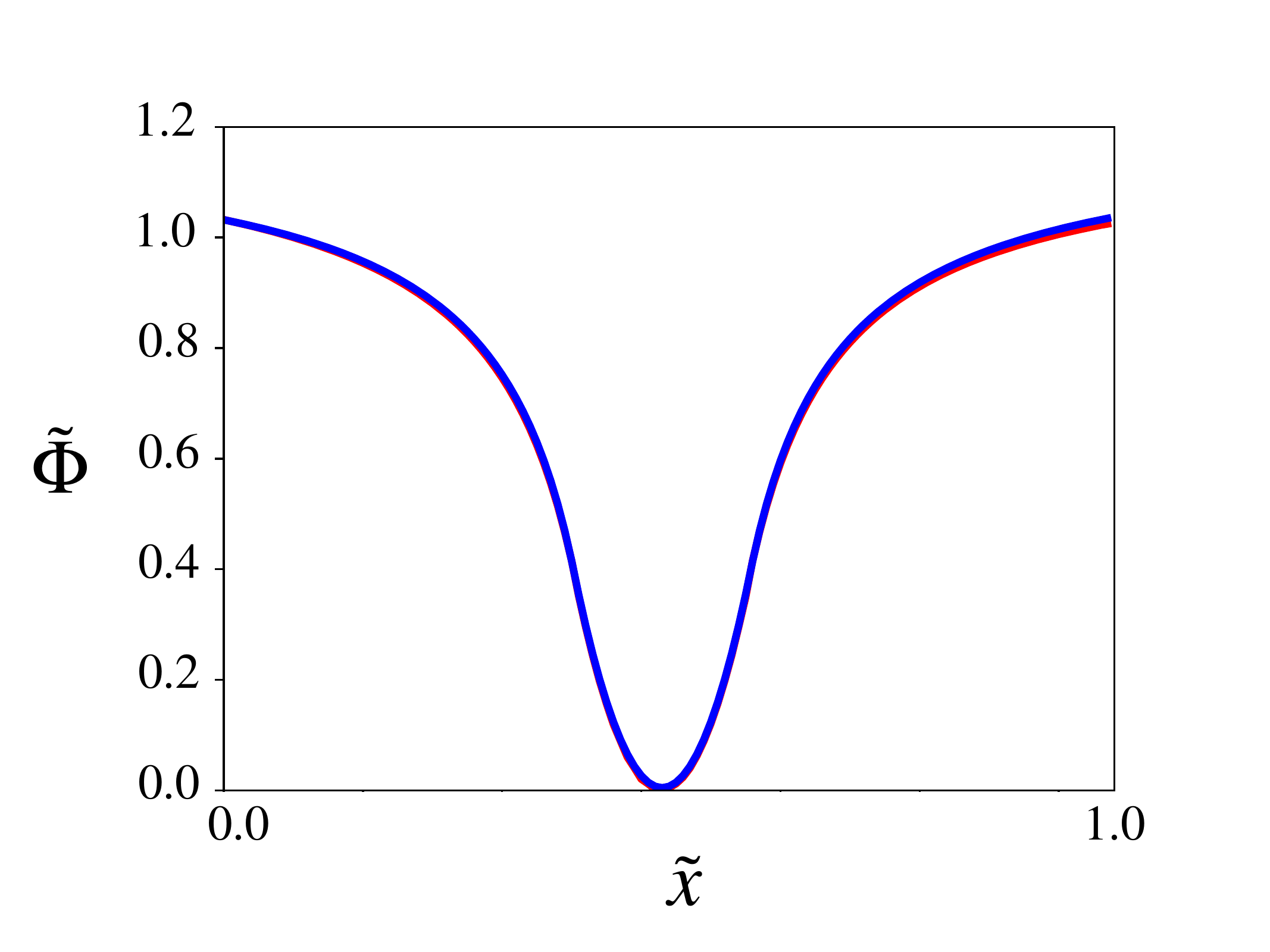}
\caption{Numerical and analytic solutions of $\tilde{\Phi}$ along the $x$-axis $(\tilde{y} = 0.5,\tilde{z} = 0.5)$ at $\tilde{t}=1.56$. The blue line indicates the numerical solution, and the red line indicates the analytic solution. \label{fig:phiexa}}
\end{figure}
 As the first test, we set a uniform density sphere with non-dimensional radius $\tilde{r}=r/L=0.1$ and density $\tilde{\rho}=10^3$, located at the center of the numerical box.
The sphere is located in a vacuum.
The resolution of this test is set to $128^3$ cells.

Figure \ref{fig:time-evl} illustrates the snapshots of the gravitational potential $\tilde{\Phi}$ (top panels) and $\Delta_\mathrm{err}$ (bottom panels) at $\tilde{z}=0.5$ plane for $\tilde{\kappa}=2.5$. 
Figure \ref{fig:phiexa} illustrates the numerical and analytic solutions of $\tilde{\Phi}$ along the $x$-axis at $(\tilde{y},\tilde{z})=(0.5,0.5)$ and $\tilde{t}=1.56$, which indicates that $\tilde{\Phi}$ already relaxes to the solution of the Poisson equation at this epoch.

                                                                                                                                                                                                                                                                  \begin{figure}
 \centering
 \includegraphics[width=0.4\textwidth]{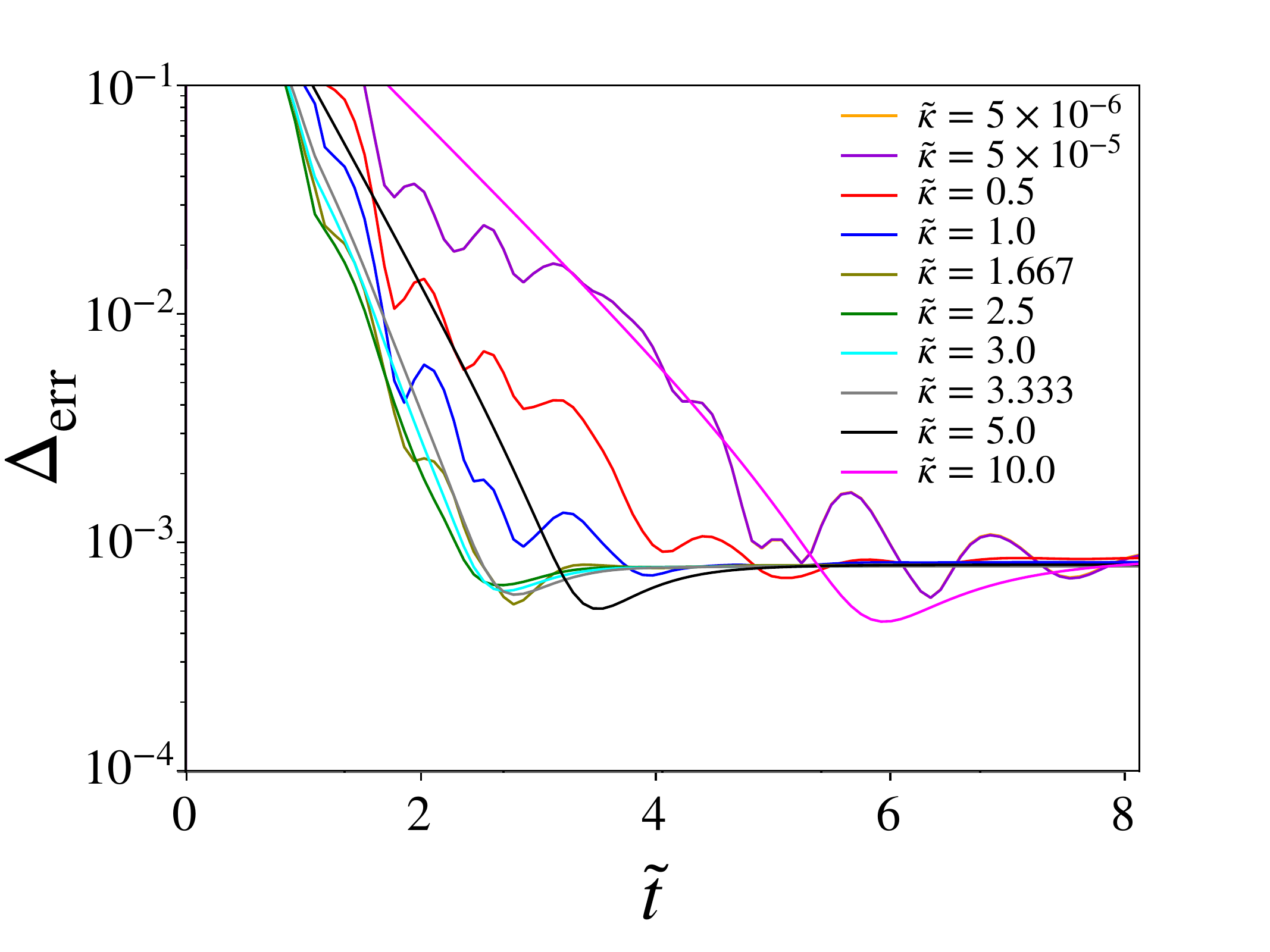}
\caption{Evolution of $\Delta_\mathrm{ err }$ for a uniform density sphere with the $\tilde{r}=0.1$ case ($128^3\ \mathrm{cells}$).
Lines of different colors correspond to the results of different choices of $\tilde{\kappa}$: $5.0\times10^{-6}$ (orange), $5.0\times10^{-5}$ (purple), $0.5$ (red), $1.0$ (blue), $1.667$ (yellow-green), $2.5$ (green), $3.0$ (cyan), $3.333$ (gray), $5.0$ (black), and $10.0$ (magenta). \label{fig:sph-r20}}
\end{figure}

Figure \ref{fig:sph-r20} illustrates the evolution of $\Delta_\mathrm{ err }$. 
Different colors correspond to the results of different choices of $\tilde{\kappa}$: $5.0\times10^{-6}$ (orange), $5.0\times10^{-5}$ (purple), $0.5$ (red), $1.0$ (blue), $1.667$ (yellow-green), $2.5$ (green), $3.0$ (cyan), $3.333$ (gray), $5.0$ (black), and $10.0$ (magenta).
$\tilde{\Phi}$ converges most rapidly in the case of $\tilde{\kappa}=2.5$. 
The reason for this can be explained as follows. 
The potential field shows oscillation due to the small damping effect of wave propagation under small $\tilde{\kappa}$. Such an oscillatory behavior can be seen in the evolution of $\Delta_\mathrm{ err }$ for the cases of $\tilde{\kappa}=5.0\times10^{-6}\ \mathrm{(orange)}$ and $5.0\times10^{-5}\ \mathrm{(purple)}$ (Fig.~\ref{fig:sph-r20}). However, under large $\tilde{\kappa}$, propagation of a gravitational wave of $\tilde{\Phi}$ is damped before the relaxation. Therefore, there is an appropriate choice of $\tilde{\kappa}$ optimized for convergence.

\begin{figure}
 \centering
 \includegraphics[width=0.4\textwidth]{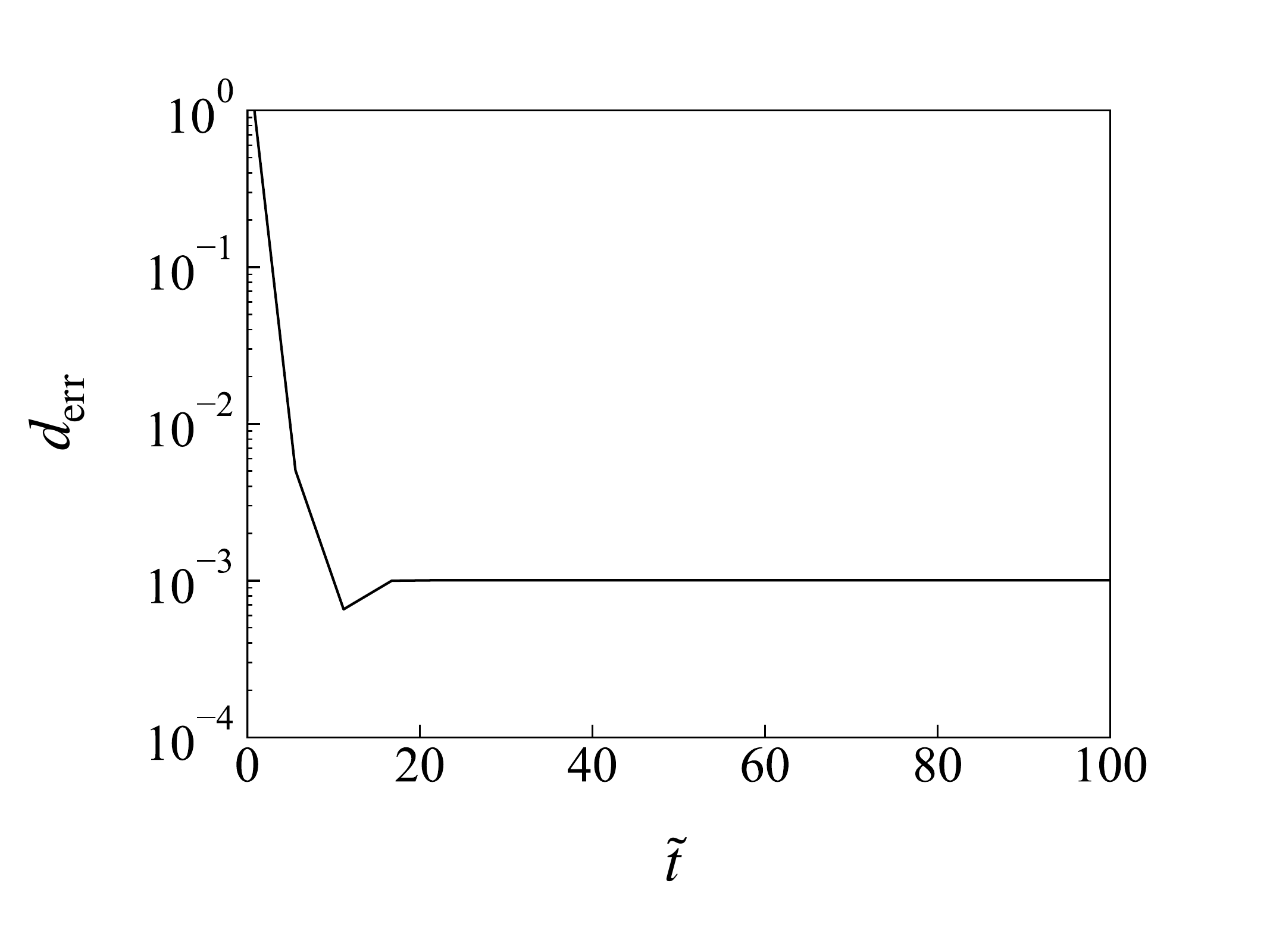}
\caption{$d_\mathrm{ err }$ evolution for a uniform density sphere with the $\tilde{r}=0.1$ case ($128^3\ \mathrm{cells}$). 
$\tilde{\kappa}$ is set to $2.5$ in this calculation.
\label{fig:tel_defect}}
\end{figure}
In addition to the error from the analytic solution, we also measure the residual error for the discretized Poisson equation as follows:
\begin{equation}
d_\mathrm{err}=\frac{\sqrt{\frac{1}{n_x n_y n_z} \Sigma_{i,j,k=1}^{n_x,n_y,n_z} \left( d_{i,j,k} \right)^{2}}}{\tilde{\Phi}_{\mathrm{exa,max }}-\tilde{\Phi}_{\mathrm{exa,min}}}, 
\end{equation}
where 
\begin{equation}
\begin{aligned} d_{i, j, k}= & \tilde{\rho}_{i, j, k}\ h^2 -\left(\tilde{\Phi}_{i-1, j, k}+\tilde{\Phi}_{i+1, j, k}+\tilde{\Phi}_{i, j-1, k}\right. \\ & \left.+\tilde{\Phi}_{i, j+1, k}+\tilde{\Phi}_{i, j, k-1}+\tilde{\Phi}_{i, j, k+1}-6 \tilde{\Phi}_{i, j, k}\right). \end{aligned}
\end{equation}
Here, $h$ in this equation is the cell size. 
The initial and boundary conditions for this calculation are the same as simulations shown in Figure \ref{fig:sph-r20}.  
The resolution is set to $128^3$ cells and $\tilde{\kappa}$ is fixed to be $2.5$. 
Figure \ref{fig:tel_defect} shows the evolution of $d_\mathrm{err}$ that indicates a similar error level to $\Delta_\mathrm{err}$. 
In the case of the multigrid method, $d_\mathrm{err}$ can be as small as a round-off error \citep[e.g.,][]{tomida2023athena++}. 
This difference would stem from the fact that our scheme does not exactly solve the Poisson equation, while the multigrid method performs iterations aiming to directly reduce $d_\mathrm{err}$. Note that the larger $d_\mathrm{err}$ of our method does not mean a larger numerical error than the multi-grid method, because the $\Delta_\mathrm{err}$, which is most important for scientific purposes, are comparable in both our method and the multigrid method.

\begin{figure}
 \centering
 \includegraphics[width=0.4\textwidth]{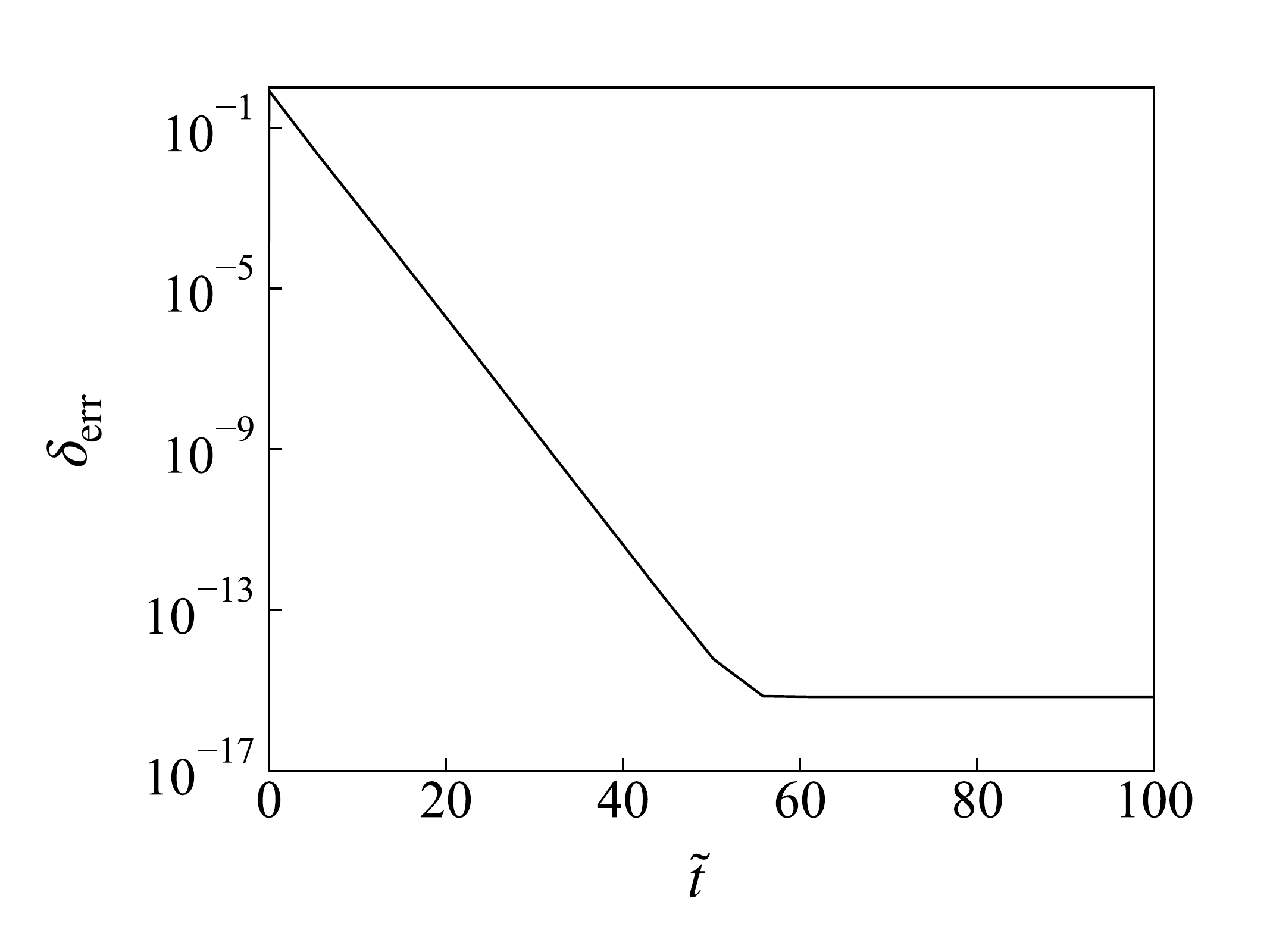}
\caption{Evolution of $\delta_\mathrm{ err }$ for a uniform density sphere with the $\tilde{r}=0.1$ case ($128^3\ \mathrm{cells}$). 
$\tilde{\kappa}$ is set to $2.5$ in this calculation. 
We use a potential after convergence as $\tilde{\Phi}_{\mathrm{cnv}}=\tilde{\Phi} \left(\tilde{t}=150 \right)$.
\label{fig:tel_disc}}
\end{figure}
We also check the error ($\delta_\mathrm{err}$) with respect to a potential after convergence, $\tilde{\Phi}_{\mathrm{cnv}}=\tilde{\Phi} \left(\tilde{t}=150 \right)$ (see Figure \ref{fig:tel_disc}). 
The definition of $\delta_\mathrm{err}$ is as follows: 
\begin{equation}
\delta_\mathrm{err}=\frac{\sqrt{\frac{1}{n_x n_y n_z} \Sigma_{i,j,k=1}^{n_x,n_y,n_z} \left( \tilde{\Phi}_{i,j,k}-\tilde{\Phi}_{\mathrm{cnv},i,j,k} \right)^{2}}}{\tilde{\Phi}_{\mathrm{cnv,max }}-\tilde{\Phi}_{\mathrm{cnv,min}}},
\end{equation}
where $\tilde{\Phi}_\mathrm{cnv, max}$ and $\tilde{\Phi}_\mathrm{cnv, min}$ represent the maximum and minimum value of the converged gravitational potential in the numerical domain. 
In this simulation, we also use a uniform density sphere with radius $\tilde{r}=0.1$ as the initial condition, the analytical solution as the boundary condition, the $128^3$ cells as the resolution of the simulation, and $\tilde{\kappa}=2.5$ as the fixed diffusion coefficient.
The result is shown in Figure \ref{fig:tel_disc}. 
In this case, the error $\delta_\mathrm{err}$ has converged to about the round-off error as expected.

\begin{figure}
 \centering
 \includegraphics[width=0.4\textwidth]{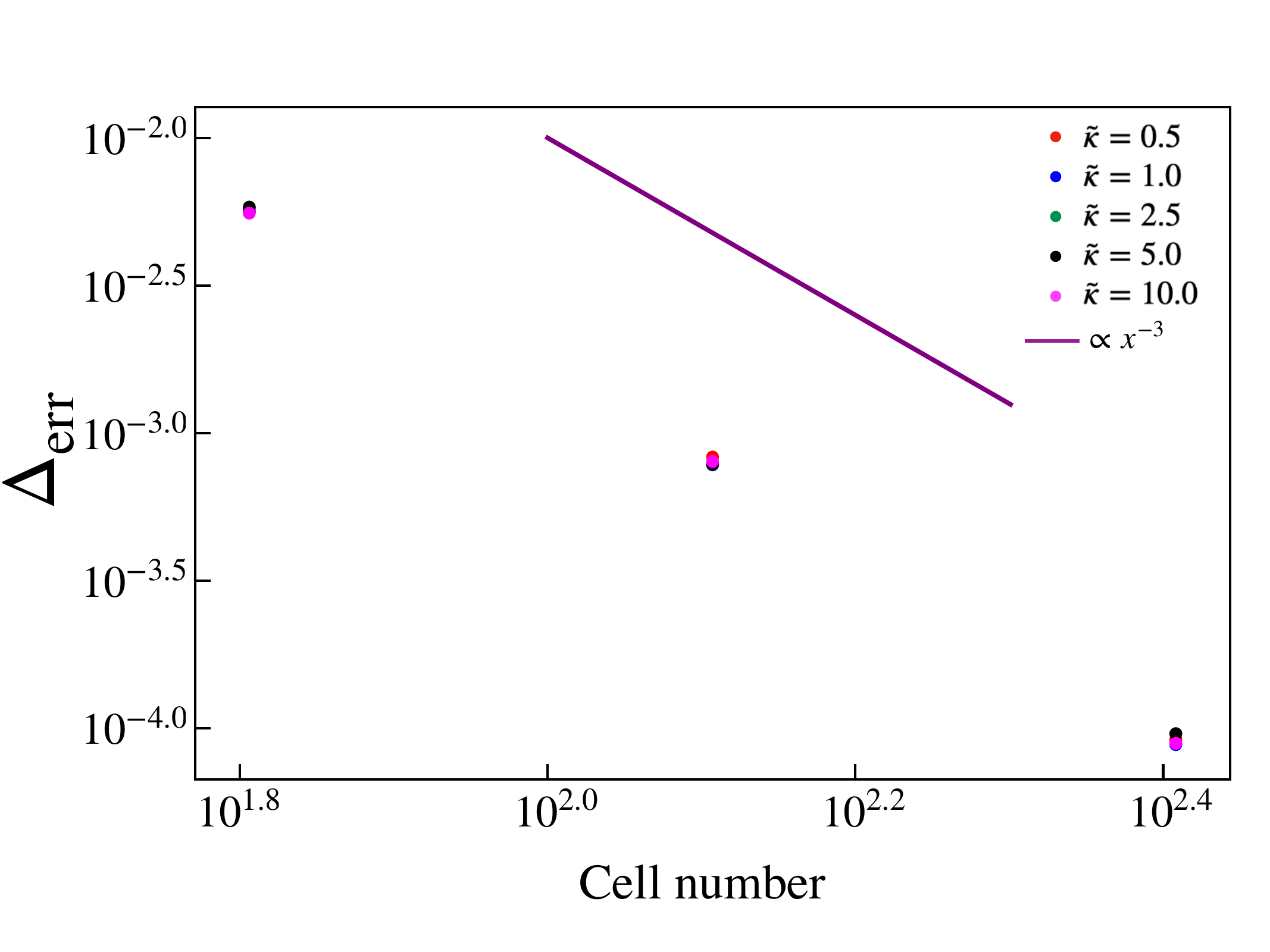}
\caption{Resolution dependence of converged $\Delta_\mathrm{ err }$ in the fiducial model. The points in the figure correspond to the results of $64^3,\ 128^3,$ and $256^3\ \mathrm{cells}$. The colors of the dots, red, blue, green, black, and magenta, correspond to the results for $\tilde{\kappa}= 0.5, 1.0, 2.5, 5.0,$ and $10.0$, respectively. 
The purple line indicates the line with a slope of $-3$.\label{fig:err}}
\end{figure}

Figure \ref{fig:err} illustrates the resolution dependence of the converged $\Delta_\mathrm{ err }$, where different colors represent the results of the different values of $\tilde{\kappa}$ (red: 0.5, blue: 1.0, green: 2.5, black: 5.0, and 
magenta: 10.0), and a solid line indicates a reference with a slope of $-3$.
Here, we plot the value of $\Delta_\mathrm{ err }$ after reaching the static state of $\tilde{\Phi}$. 
We can confirm that $\Delta_\mathrm{ err }$ decreases as it is designed (third-order spatial accuracy), and the converged solution does not depend on the choice of $\tilde{\kappa}$.

\subsubsection{Results for various density distributions}

The suitable value of $\tilde{\kappa}$ could vary with the density distribution since the parameters of the density distribution is a crucial factor for the solution of $\tilde{\Phi}$.
In the following, we examine whether the appropriate value of $\tilde{\kappa}$ for convergence varies with the density distribution. 
In the following computations, the total mass in the numerical domain is the same as the fiducial model, and the resolution is fixed to $128^3$ cells.

\begin{figure}
 \centering
 \includegraphics[width=0.4\textwidth]{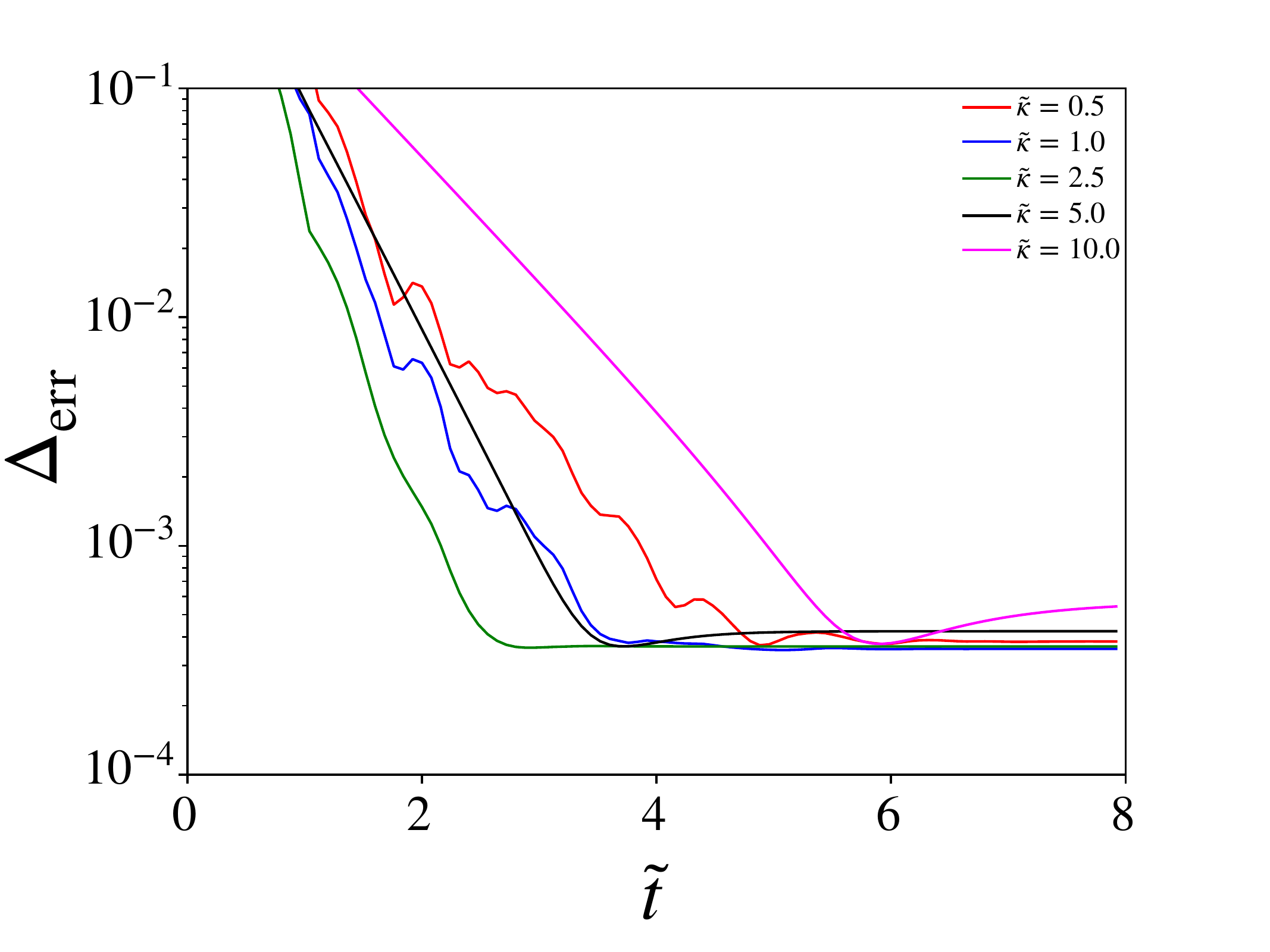}
\caption{Evolution of $\Delta_\mathrm{ err }$ for a uniform density sphere with the $\tilde{r}=0.45$ case ($128^3\ \mathrm{cells}$).
Lines of different colors show the results of $\tilde{\kappa}=0.5$ (red), $1.0$ (blue), $2.5$ (green), $5.0$ (black), and $10.0$ (magenta).
\label{fig:sph-r45-err}}
\end{figure}

\begin{figure}
 \centering
 \includegraphics[width=0.4\textwidth]{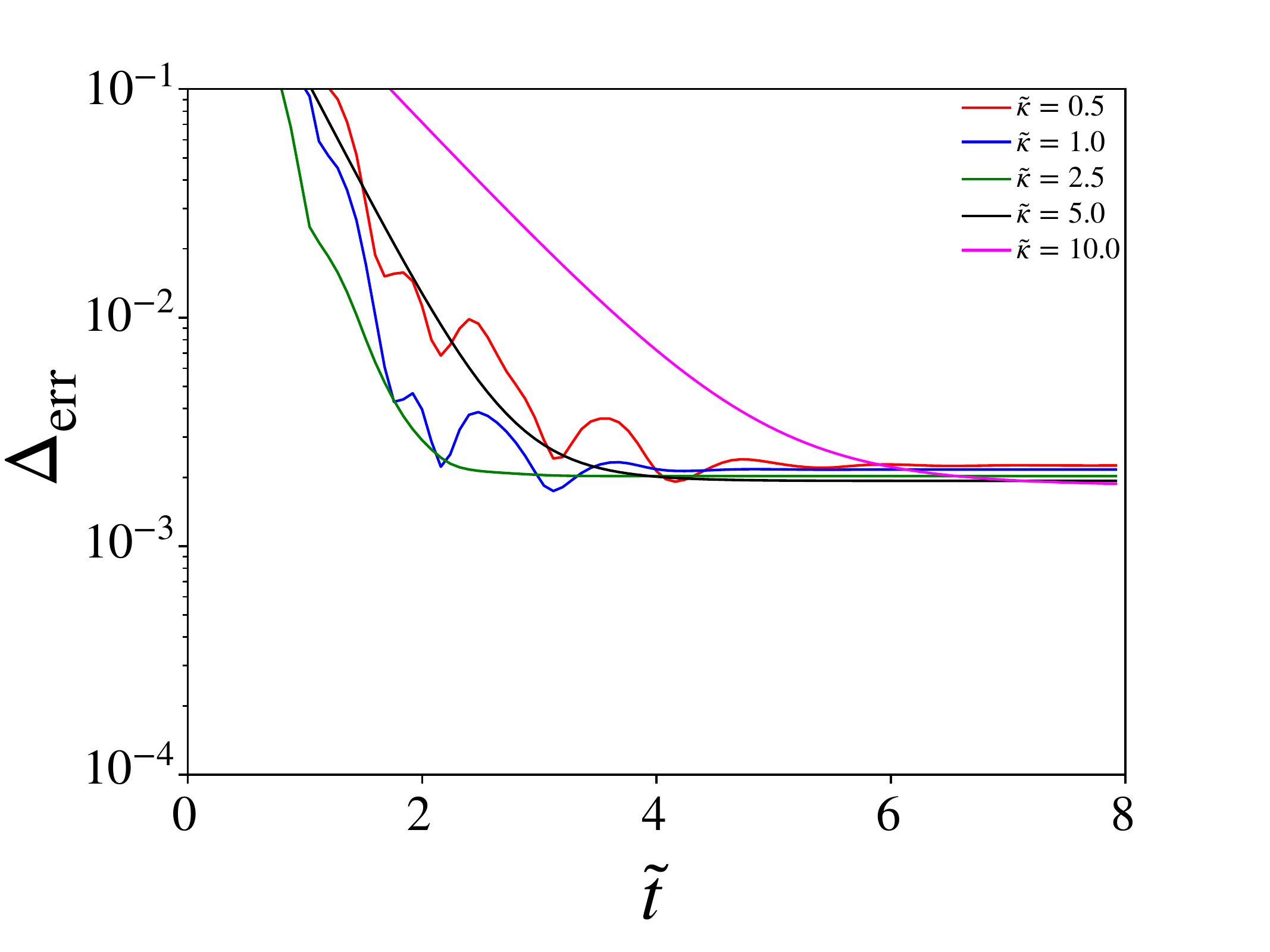}
\caption{Evolution of $\Delta_\mathrm{ err }$ for a uniform density sphere with the $\tilde{r}=0.02$ case ($128^3\ \mathrm{cells}$). 
Lines of different colors indicate the results of $\tilde{\kappa}=0.5$ (red), $1.0$ (blue), $2.5$ (green), $5.0$ (black), and $10.0$ (magenta).
\label{fig:sph-r002-err}}
\end{figure}

We change a radius of a sphere from the previous model and set $\tilde{r}=0.45$ and $\tilde{r}=0.02$.
Figure \ref{fig:sph-r45-err} ($\tilde{r}=0.45$) and Figure \ref{fig:sph-r002-err} ($\tilde{r}=0.02$) show the results of these tests. The results clearly show that $\tilde{\kappa}=2.5$ leads to the fastest convergence.

\begin{figure}
 \centering
 \includegraphics[width=0.4\textwidth]{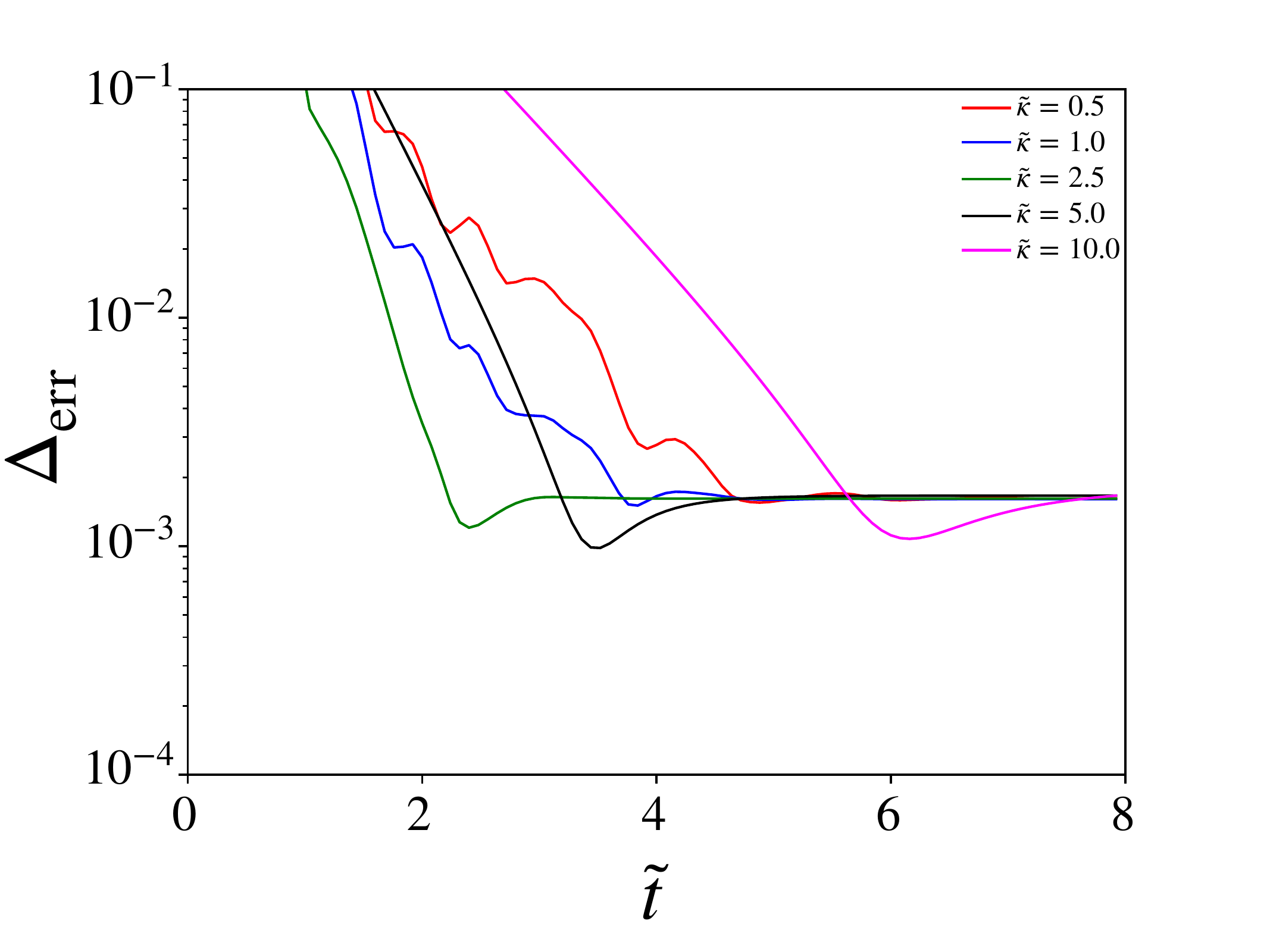}
\caption{Evolution of $\Delta_\mathrm{ err }$ for the dipole density distribution case ($128^3\ \mathrm{cells}$). 
Lines of different colors correspond to the results of $\tilde{\kappa}=0.5$ (red), $1.0$ (blue), $2.5$ (green), $5.0$ (black), and $10.0$ (magenta).
\label{fig:dipole-err}}
\end{figure}

\begin{figure}
 \centering
 \includegraphics[width=0.4\textwidth]{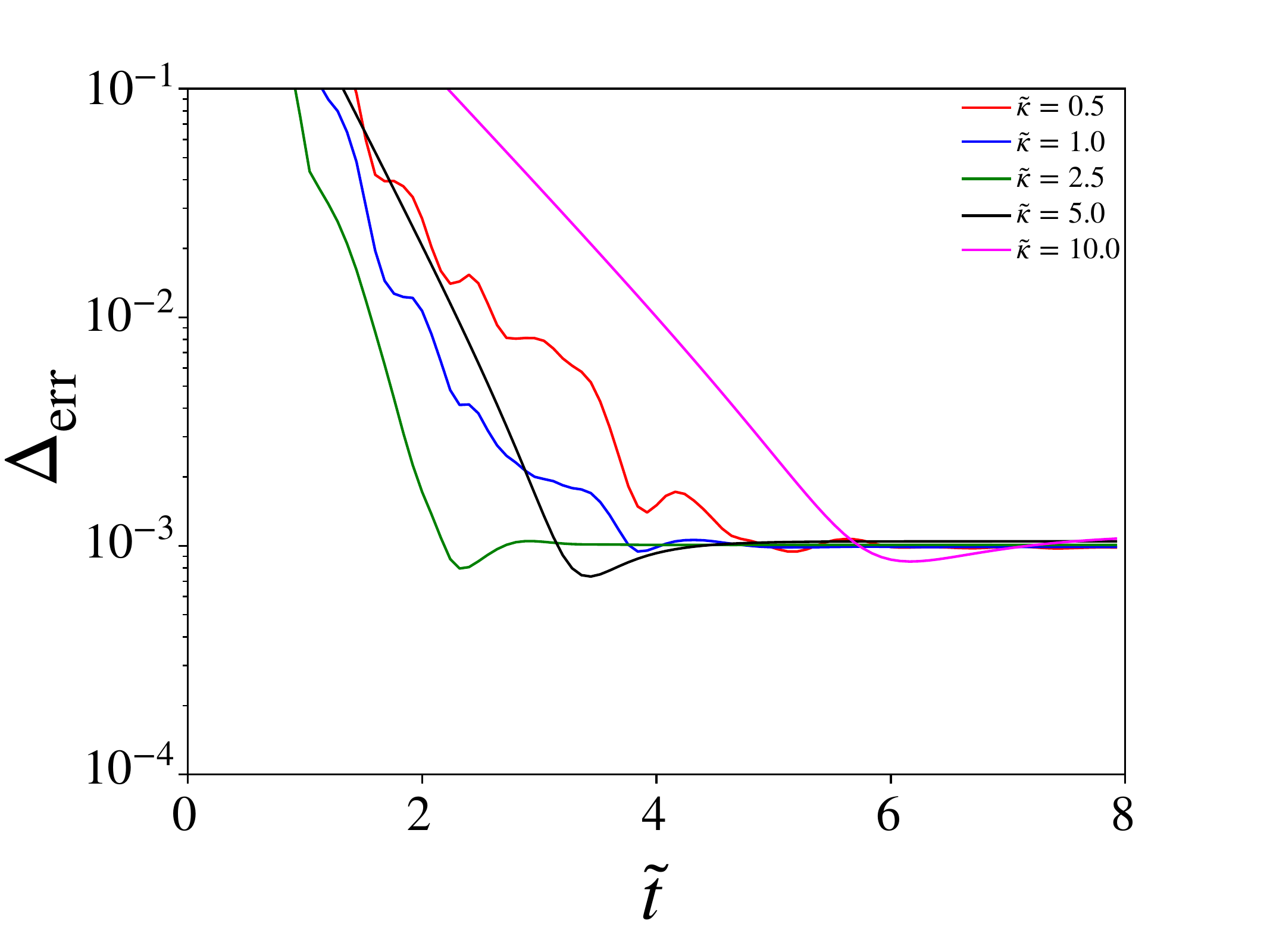}
\caption{Evolution of $\Delta_\mathrm{ err }$ for the quadrupole density distribution case ($128^3\ \mathrm{cells}$). 
Lines of different colors show the results of $\tilde{\kappa}=0.5$ (red), $1.0$ (blue), $2.5$ (green), $5.0$ (black), and $10.0$ (magenta).
\label{fig:quad-err}}
\end{figure}

We also analyze the cases of dipole and quadrupole density distributions. 
In the dipole case, two uniform density spheres with $\tilde{r}=0.1$ and $\tilde{\rho}=5.0\times 10^2$ are set at $(\tilde{x},\tilde{y},\tilde{z})=(0.25,0.25,0.25)$ and $(0.75,0.75,0.75)$. 
In the quadrupole case, we employ four uniform density spheres with $\tilde{r}=0.1$ and $\tilde{\rho}=2.5\times 10^2$ located at $(\tilde{x},\tilde{y},\tilde{z})=(0.25,0.25,0.25), (0.25,0.75,0.75), (0.75,0.25,0.25)$, and $(0.75,0.75,0.75)$. 
The evolution of $\Delta_\mathrm{ err }$ for dipole and quadrupole density distributions are illustrated in Figures \ref{fig:dipole-err} and \ref{fig:quad-err}, respectively.
Different colors represent the results of different choices of $\tilde{\kappa}$ (red: 0.5, blue: 1.0, green: 2.5, black: 5.0, and 
magenta: 10.0). 
The results again indicate that $\tilde{\kappa}=2.5$ gives the best performance irrespective of the density distribution.

\subsection{Case of a moving density field}
In this section, we analyze the case of a time-dependent density field, particularly a rotating system. From these tests, we can derive a necessary condition on the choice of $c_\mathrm{g}$ value for convergence.

We consider a uniform density sphere which circulates in a numerical domain with velocity $\tilde{v}_\mathrm{c}$ and orbital radius $\tilde{r}_\mathrm{c}$. 
Here, $\tilde{v}_\mathrm{c}$ represents a dimensionless velocity given by $\tilde{v}_\mathrm{c}=v_\mathrm{c}/c_\mathrm{g}$.
The sphere’s density and radius are $\tilde{\rho}=10^3$ and $\tilde{r}=0.1$, respectively.
We set the center of the circular motion as the center of the simulation box and $\tilde{r}_\mathrm{c}= 0.35$.
We analyze three cases with different velocities of the circular motion, $\tilde{v}_\mathrm{c}=1.0, 0.1,$ and $0.01$.
The analytic solution for the initial gravitational potential is employed as an initial condition for $\tilde{\Phi}$. 
The resolution is set to $128^3$ cells and the simulations are performed using $\tilde{\kappa}=2.5$. 

\begin{figure}
 \centering
 \includegraphics[width=0.4\textwidth]{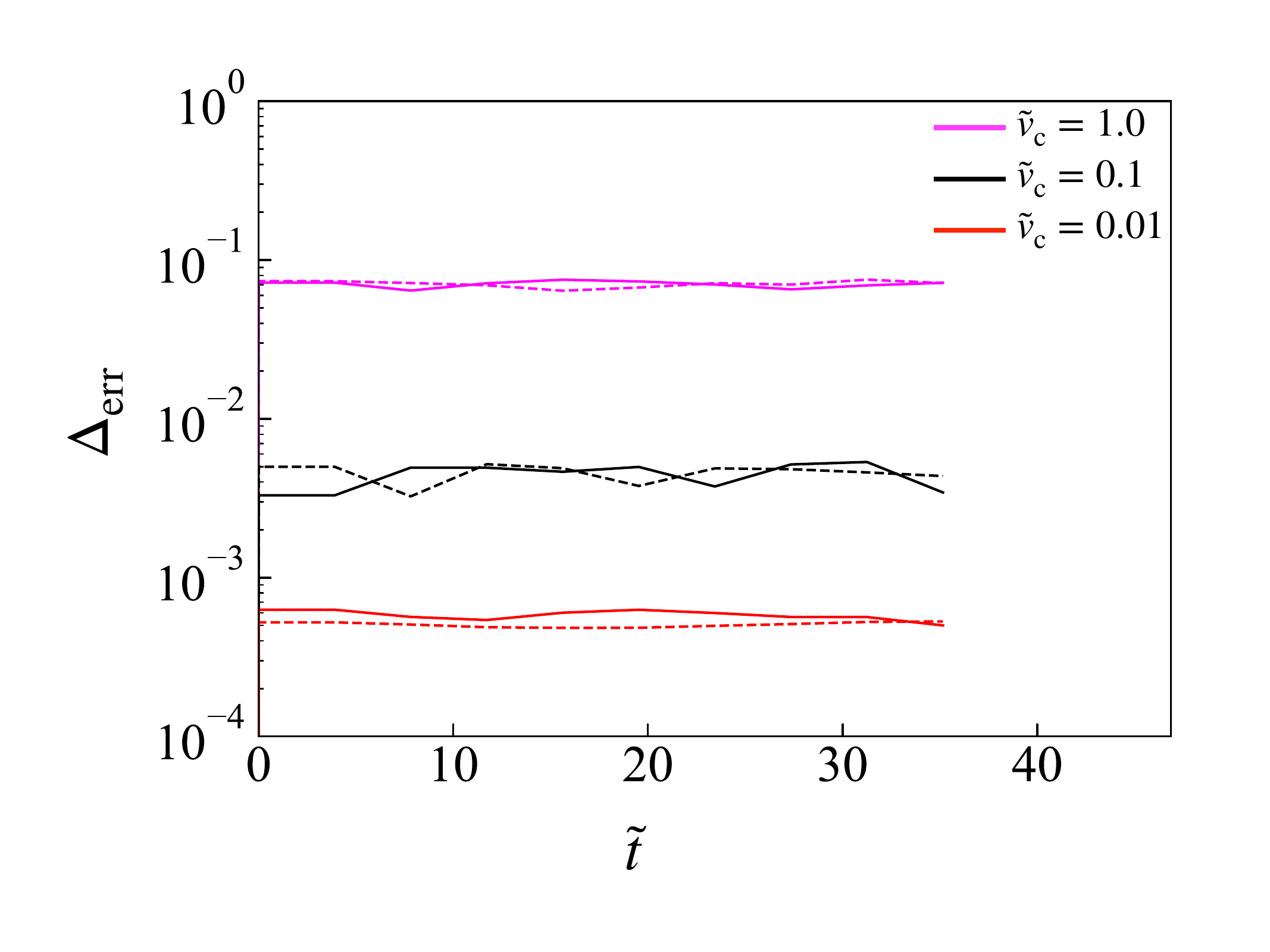}
\caption{Evolution of $\Delta_\mathrm{ err }$ when a uniform density sphere with $\tilde{r} = 0.1$ moves in a circular motion with velocity $\tilde{v}_\mathrm{c}$. 
Lines of different colors indicate the results of $\tilde{v_\mathrm{c}}=0.01$ (red), $0.1$ (black), and $1.0$ (magenta).
The resolution is set to  $128^3$ (solid lines), $256^3$ (dashed lines) cells, and the simulation is performed for $\tilde{\kappa}=2.5$. 
\label{fig:tel-mv}}
\end{figure}

Figure \ref{fig:tel-mv} illustrates the resulting $\Delta_\mathrm{ err }$ evolution for the rotating density field.
The magenta, black, and red lines correspond to results of $\tilde{v}_\mathrm{c}=1.0, 0.1,$ and $0.01$, respectively. 
The resolution is set to $128^3$ (solid lines), $256^3$ (dashed lines). 
This figure shows that if $v_\mathrm{c}$ is taken to be sufficiently small compared to $c_\mathrm{g}$, $\Delta_\mathrm{ err }$ is kept at a small value. 
The error (differences between the solutions of the Poisson equation and eq.~(\ref{eq:tel})) decreases as $\Delta_\mathrm{err} \propto (v_\mathrm{c}/c_\mathrm{g})^2$, according to \cite{hirai2016hyperbolic}. 
This is simply because gravitational potential can relax sufficiently before the density field changes. 
$\Delta_\mathrm{ err }$ is suppressed to be less than 1\% if we select as $v_\mathrm{c}/c_\mathrm{g} \lesssim 0.1$  regardless of the resolution.
Therefore, we conclude that the numerical error can be suppressed, even if we employ a small value of $c_\mathrm{g}$ instead of $c$. The detailed choice of $c_\mathrm{g}$ depends on prescribed accuracy, but a value of $c_\mathrm{g}$ ten times larger than gas velocity makes the error smaller than 1\%.
 
\begin{figure}
 \centering
 \includegraphics[width=0.4\textwidth]{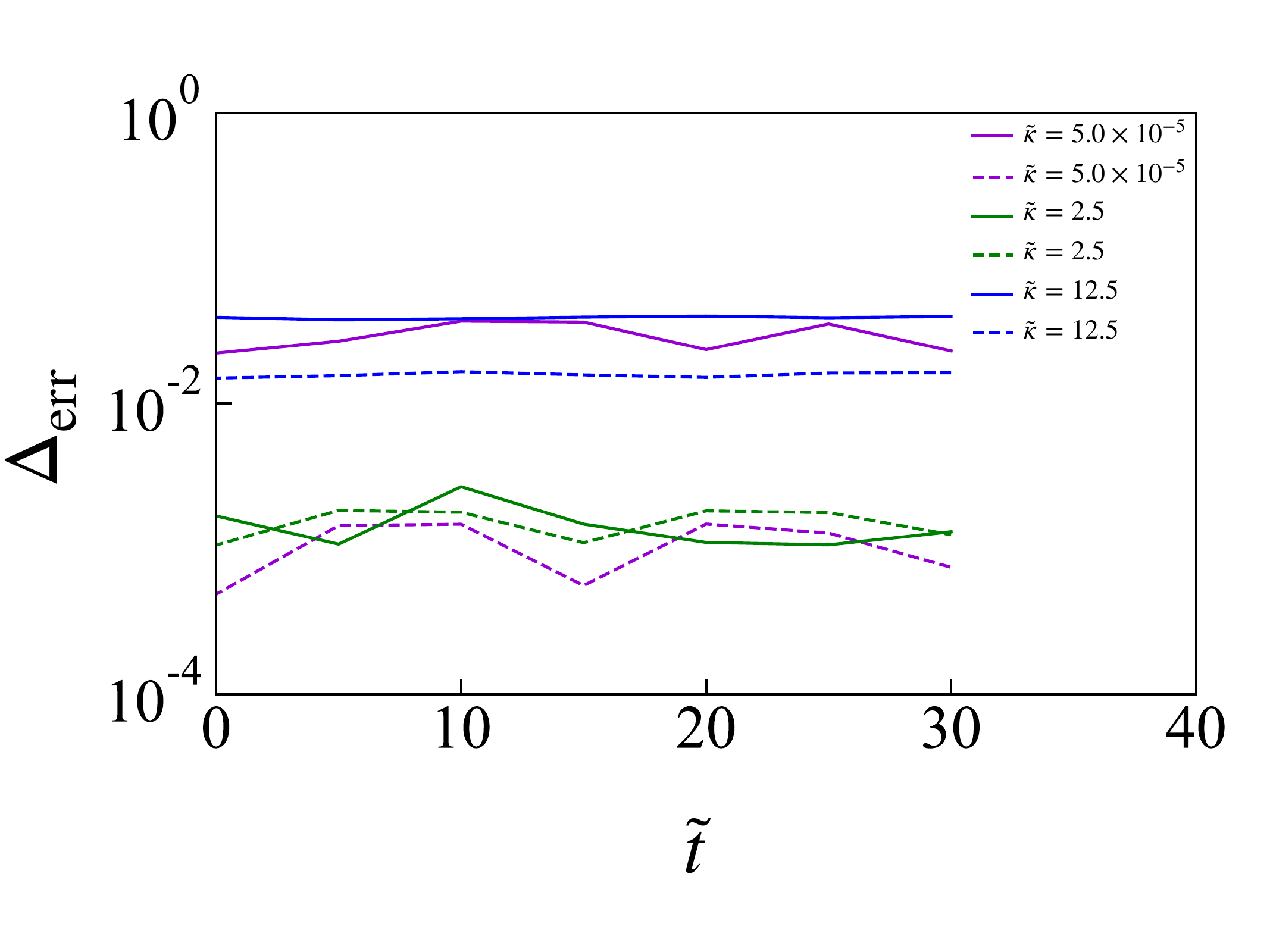}
\caption{Evolution of $\Delta_\mathrm{ err }$ when a uniform density sphere with $\tilde{r} = 0.1$ moves in a circular motion with velocity $\tilde{v}_\mathrm{c}=0.1$. 
The colors show the choice of $\tilde{\kappa}$ (purple: $5\times10^{-5}$, green: 2.5, and blue: 12.5). 
The dashed (solid) lines mean the results of fixed boundary conditions (periodic boundary conditions). 
\label{fig:tel-cyc-prd}}
\end{figure}

Furthermore, we examine the influence of the choice of $\tilde{\kappa}$ on $\Delta_\mathrm{ err }$ in the rotating system.
Figure \ref{fig:tel-cyc-prd} illustrates the resulting $\Delta_\mathrm{ err }$ for different values of $\tilde{\kappa}$ under the condition of $\tilde{v}_\mathrm{c}=0.1$.  Colors represent the choice of $\tilde{\kappa}$ (purple: $5\times10^{-5}$, green: 2.5, and blue: 12.5).
The line type shows the choice of the boundary conditions: solid lines indicate the results for the periodic boundary and dashed lines for the exact boundary condition mentioned in Section 2.2. 
In the case of periodic boundary conditions, it is difficult to obtain the analytic solution for $\tilde{\Phi}$; thus, the well-converged $\tilde{\Phi}$ is used instead of $\tilde{\Phi}_\mathrm{exa}$ in eq. (\ref{eq:err}) to compute $\Delta_\mathrm{err}$. 
Figure \ref{fig:tel-cyc-prd} demonstrates that the numerical error can be suppressed smaller than 1\% independent of the choice of boundary conditions in the case of $\tilde{\kappa}=2.5$. 
Particularly, for the periodic boundary condition, the error is kept small compared to the $\tilde{\kappa}=5.0\times 10^{-5},$ and $12.5$ cases. 
In the this system, the deviations of $\tilde{\Phi}$ from the converged solution propagate as waves.
In the case of periodic boundary conditions, such waves do not go outside the numerical domain and travel in the computational domain long time as errors. 
Therefore, suitable damping is crucial to make convergence faster. 
These results indicate that $\tilde{\kappa}=2.5$ offers the best performance even for a time-dependent density field.

\section{discussion}
In this section, we compare our method with other self-gravity calculation methods. 
We conducted weak-scaling tests to verify the parallel efficiency performance. 
Our tests were carried out on the supercomputer "Flow" at Information Technology Center, Nagoya University. 
Here we compare our results with other methods shown by \cite{tomida2023athena++}, which was conducted using Cray XC50 supercomputer at Center for Computational Astrophysics, National Astronomical Observatory of Japan. 
In the following, we show the results where each process has $64^3$ cells.

\begin{figure}
 \centering
 \includegraphics[width=0.45\textwidth]{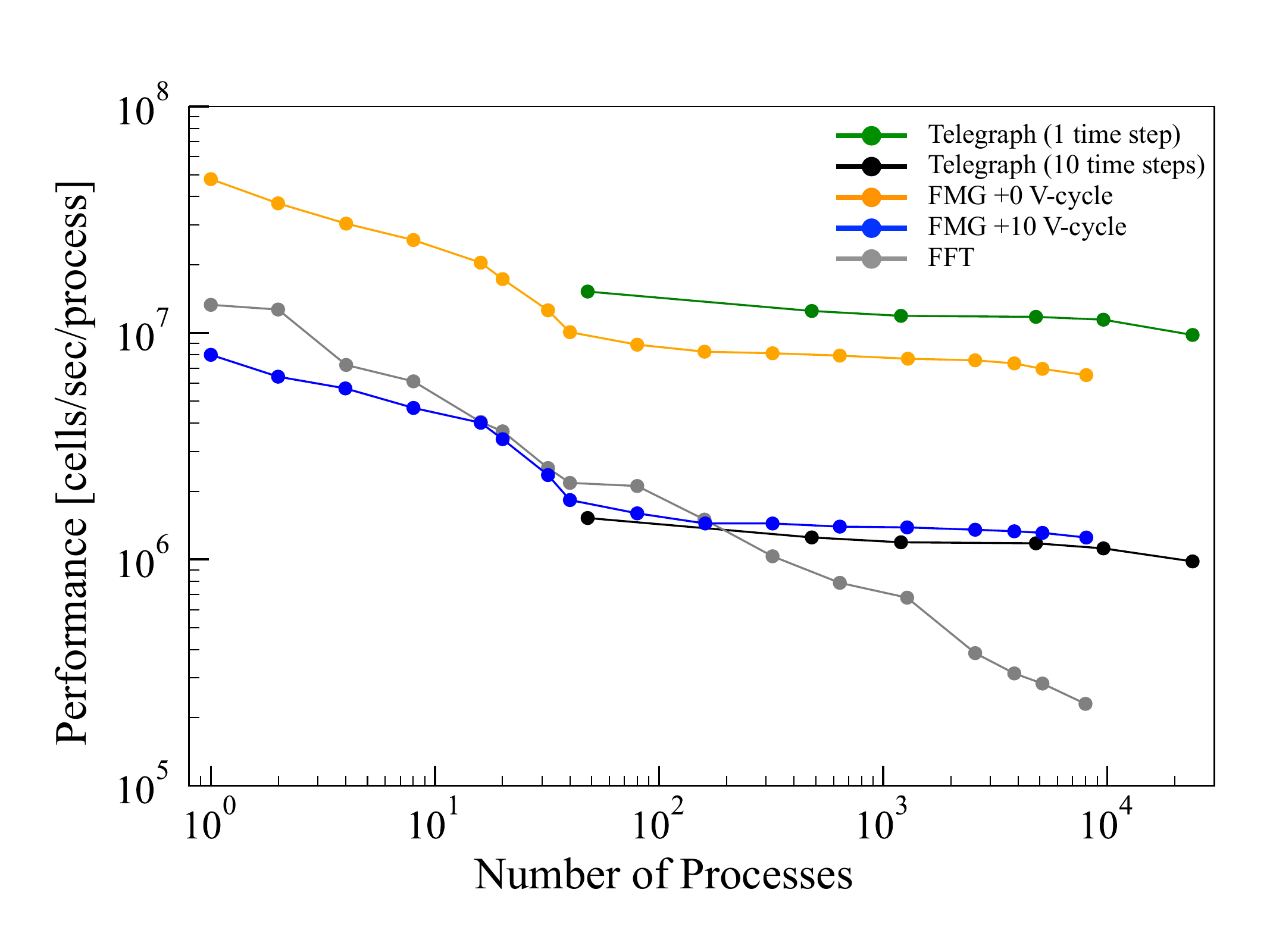}
\caption{Results of weak-scaling tests. The vertical and horizontal axes represent performance and number of processes, respectively. 
The colors represent the results of the telegraph method with 1 time-step (green), the telegraph method with 10 time-step (black), the FMG method with 0 V-cycle (orange), the FMG method with 10 V-cycle (blue), and FFT. In all calculations, one process has $64^3$ cells. 
Results of FMG and FFT methods are performed in Tomida \& Stone (2023). 
\label{fig:perf}}
\end{figure}
 
The green and black lines in Figure \ref{fig:perf} represent the 1 and 10 time step performance of our telegraph solver, respectively. 
If we choose $c_\mathrm{g}=10\ v_\mathrm{c}$, we need 10 inner time steps to integrate eq.(\ref{eq:tel}) and to obtain the potential $\Phi$. 
Each point is the result of the $48$, $480$, $1200$, $4800$, $9600$, and $24000$ processes calculation. 
We averaged the computation time over 4000 time steps to obtain an accurate value of performance. 
We can confirm that our method maintains high performance even for $>10^4$ parallel calculations, thanks to the hyperbolic nature of our basic equations. 
 Note that we did not care about the arrangement of the nodes in our calculations, and this would be the reason why the performance decreased slightly for the process number $\gtrsim 10^4$. 
Note also that we are computing with flat MPI, while the OpenMPI/MPI hybrid computation is employed in \cite{tomida2023athena++}.

The gray line in Figure \ref{fig:perf} represents the performance of the FFT self-gravity calculation shown in \cite{tomida2023athena++}. 
Because FFT is computationally expensive for large-scale calculations, its performance decreases for massive parallel calculations. 

The orange and blue lines in Figure \ref{fig:perf} are the results of the full multigrid (FMG) method obtained in \cite{tomida2023athena++}. 
The orange line shows the result of a single FMG calculation, and the blue line shows the result of the FMG plus 10 V-cycles calculation. 
Note that these results are averages of the best 10 out of 100 runs. 
The figure shows that the FMG method has as high parallel efficiency as our method for parallel computations up to about $10^4$ processes. 
However, when considering more large parallel computations, the multigrid method would decrease the efficiency due to the parallelization difficulty of the elliptic equation. 
In addition, our method can be very useful for massive parallel computations in particular when grid structure is inhomogeneous like adaptive mesh refinement (AMR) simulations, which substantially impairs parallelization in the multigrid method.

The multigrid methods need several times of iterations for convergence. For instance, in the SFUMATO code \citep{matsumoto2007self}, three iterations (V-cycle) are imposed to calculate the potential $\Phi$. 
However, the FMG method has been shown to reduce errors even when no iterations are performed (one cycle of FMG costs about two V-cycles). 
If our method requires 10 iterations, then the performance will be less than the FMG method, which requires 0 iterations in the range of up to $10^4$ process calculations. 
However, in the range of $>10^4$, where FMG's parallelization efficiency would decrease, we expect our method to be superior to FMG's performance.

More importantly, our method shows much higher performance than the FMG method when the time step of the simulation is determined by cooling time, chemical reactions, etc. ($\Delta t_\mathrm{c, cr, \cdots}$), rather than the CFL condition for the fluid ($\Delta t_\mathrm{f}$).
As shown in Figure \ref{fig:tel-cyc-prd}, our method requires the gravitational field to evolve more rapidly than the fluid as in $c_\mathrm{g}=N\ (c_\mathrm{s}+v_\mathrm{fluid})$, where $N\gg 1$. 
Here, our self-gravity solver must satisfy the CFL condition of $c_\mathrm{g}$ to evolve $\tilde{\Phi}$ and $\tilde{\Psi}$. 
Thus, when the time step of the original calculation is determined by the CFL condition for the fluid $\Delta t =\Delta t_\mathrm{f}=0.5\Delta x/ (v_\mathrm{fluid}+c_\mathrm{s})$, simulations using our method must be conducted with a smaller time step determined by the $c_\mathrm{g}$ CFL condition ($\Delta t =\Delta t_\mathrm{g}=0.5\Delta x/c_\mathrm{g}=\Delta t_\mathrm{f}/N$) or with the $N$ inner time steps of self-gravity using $\Delta t_\mathrm{g}$. 
However, when the time step is limited by other physics such as radiative cooling or chemical reaction, etc., the time step can be much smaller than the CFL condition for the fluid, i.e., $\Delta t=\Delta t_\mathrm{c, cr, \cdots} < \Delta t_\mathrm{f}/N \ll \Delta t_\mathrm{f}$. 
In this case, if we set $c_\mathrm{g}$ to $c_\mathrm{g}=0.5 \Delta x / \Delta t$ to satisfy the CFL condition for the gravitational phase velocity, then $c_\mathrm{g}$ also satisfies condition $c_\mathrm{g}=0.5 \Delta x / \Delta t_\mathrm{c, cr, \cdots} \gg N\ (v_\mathrm{fluid}+c_\mathrm{s})$, which means that our method can evolve self-gravity by $\Delta t$ without iteration. 
Our method is the most efficient method when no iterations are required, as is clear from Figure \ref{fig:perf}. 
Note that $N$ should be $\gtrsim 10$ as shown in Figure \ref{fig:tel-cyc-prd}.

\section{summary}
We have developed an enhanced dynamical Poisson equation solver for self-gravity, which can be employed for massive parallel computation for gravitational hydrodynamics.
\begin{itemize}
    \item[1.] In the case of a static density, we found the best damping coefficient as $\tilde{\kappa}\simeq 2.5$ regardless of the density distribution. 
    Using $\tilde{\kappa}=2.5$, the convergence of $\tilde{\Phi}$ is twice as fast as that of the $\tilde{\kappa}\simeq 0$ case. 
    This is because the oscillations of potential waves are damped by appropriate diffusion.
    \item[2.] In the case of a moving density, the error of the gravity is suppressed by selecting a faster phase velocity of the gravitational wave $c_\mathrm{g}$ than the fluid velocity. This is consistent with the results of \cite{hirai2016hyperbolic}.
    The error is kept below 1\% using $c_\mathrm{g}$ that is 10 times larger than the velocity of fluid motion.
    \item[3.] For periodic boundary conditions, our method is useful to keep the error to less than 1\%.
    In our method, the deviations of $\tilde{\Phi}$ from the converged solution propagate as waves. Such waves do not escape from the computational domain due to  periodic boundary conditions and travel in the computational domain for a long time as errors. Therefore, we should not use very small $\kappa$ and should introduce appropriate damping to suppress the error in the case of periodic boundary conditions. 
    \item[4.]  Our method maintains high performance even for $>10^4$ parallel calculations thanks to the nature of a hyperbolic equation. 
    Therefore, our method is likely to be one of the most useful methods for massive parallel computations. 
    In particular, if the time step of the calculation is determined by heating/cooling or chemical reactions, etc., our method can significantly reduce the computational cost because $c_\mathrm{g}$ determined by the CFL condition using $\Delta t_\mathrm{c, cr, \cdots}$ is much larger than the fluid velocity plus sound speed. 
    \item[5.] Although we explain our method by employing the Poisson equation for Newtonian gravity as examples in this study, our method can be used for any type of elliptic equation in principle. 
\end{itemize}

\section*{Acknowledgements}
We thank K. Omukai and K. Tomida for their helpful comments and suggestions.
Numerical computations were conducted on Cray XC50 at the Center for Computational Astrophysics, National Astronomical Observatory of Japan. 
The computation was also carried out using the JHPCN Joint Research Projects on supercomputer "Flow" at Information Technology Center, Nagoya University.
This work is supported by grant-in-aid from the Ministry of Education, Culture, Sports, Science, 
and Technology (MEXT) of Japan, Grant No.18H05436, No.18H05437, and No.20H01944.

\section*{Data Availability}
The data underlying this article will be shared on reasonable request to the corresponding author.

\bibliographystyle{mnras}
\bibliography{mnras_template.bib}

\bsp	
\label{lastpage}
\end{document}